\documentclass[useAMS,usenatbib]{mn2e}

\usepackage{graphicx}

\title[Abundances of lithium, sodium, and potassium in Vega]
{Abundances of lithium, sodium, and potassium in Vega}
\author[Y. Takeda]{Y. Takeda\thanks{E-mail:
takeda.yoichi@nao.ac.jp}\footnotemark[0]\thanks{Based on observations
carried out at Okayama Astrophysical Observatory (Okayama, Japan).}\\
National Astronomical Observatory of Japan, \\
2-21-1 Osawa, Mitaka, Tokyo 181-8588, Japan\\
}
\begin{document}

\date{Accepted 2008 May 12. Received 2008 May 8; in original form 2007 December 14}


\maketitle

\label{firstpage}

\begin{abstract}
Vega's photospheric abundances of Li, Na, and K were determined 
by using considerably weak lines measured on the very high-S/N spectrum, 
while the non-LTE correction and the gravity-darkening correction 
were adequately taken into account. 
It was confirmed that these alkali elements are mildly 
underabundant ([Li/H] $\simeq -0.6$, [Na/H] $\simeq -0.3$, and 
[K/H] $\simeq -0.2$) compared to the solar system values, 
as generally seen also in other metals.
Since the tendency of Li being more deficient than Na and K is 
qualitatively similar to what is seen in typical interstellar 
cloud, the process of interstellar gas accretion may be related 
with the abundance anomaly of Vega, as suspected in the case of 
$\lambda$ Boo stars.
\end{abstract}


\section{Introduction}

While Vega (= $\alpha$~Lyr = HR~7001 = HD~172167 = HIP~91262; A0V)
plays an important role as the fundamental photometric standard,
its photospheric chemical composition is known to be sort of
anomalous. Namely, most elements are mildly underabundant
(by $\sim -0.5$ dex on the average) in spite of its definite population I
nature, though the extent of deficiency is inconspicuous for several
volatile elements of low condensation temperature ($T_{\rm c}$) such
as C, N, O, and S. Since this is the tendency more manifestly shown 
by a group of A--F main-sequence stars known as ``$\lambda$~Boo-type 
stars'' (see, e.g., Paunzen 2004 and the references therein), 
Vega's moderate abundance peculiarities have occasionally been argued
in connection with this $\lambda$~Boo phenomenon (e.g., Baschek \& Slettebak 1998;
St\"{u}renburg 1993; Holweger \& Rentzsch-Holm 1995; Iliji\'{c} et al. 1998).

Regarding the origin of unusual surface compositions of $\lambda$~Boo 
stars, various models have been propounded so far, such as the 
diffusion/mass-loss model, accretion/diffusion model, or binary model 
(see the references quoted in Paunzen 2004). Among these, the 
interpretation that the anomaly was built-up by the accretion of 
interstellar gas (where refractory metals of high $T_{\rm c}$ are 
depleted because of being condensed into dust while volatile species 
of low $T_{\rm c}$ hardly suffer this process), such as the interaction model 
between a star and the diffuse interstellar cloud proposed by 
Kamp \& Paunzen (2002), appears to be particularly promising,
in view of Paunzen et al.'s (2002) recent finding that the [Na/H] 
values of $\lambda$ Boo stars (showing a large diversity) are closely 
correlated with the [Na/H]$_{\rm ISM}$ values (sodium abundance of
interstellar matter) in the surrounding environment (cf. Fig.~7 therein).

Then, according to the suspected connection between Vega and $\lambda$~Boo 
stars, it is natural to examine the photospheric Na abundance of Vega.
Besides, the abundances of Li and K may also be worth particular attention 
in a similar analogy, since these three alkali elements play significant roles 
in discussing the physical state of the cool interstellar gas through their 
interstellar absorption (resonance) lines of Na~{\sc i} 5889/5896 
(D$_{1}$/D$_{2}$), Li~{\sc i} 6708, and K~{\sc i} 7665/7699. 

It is somewhat surprising, however, that reliable determinations of Vega's Na 
abundance are barely available. To our knowledge, although a number of abundance 
analyses of Vega have been published so far over the past half century, Na abundance 
derivations were done only in four studies (among which original $W_{\lambda}$ 
measurements were tried only two of these) according to our literature survey:
Hunger (1955; Na~{\sc i} 5889/5896), the reanalysis of Hunger's $W_{\lambda}$
data by Strom, Gingerich \& Strom (1966),  Qiu et al. (2001; Na~{\sc i} 5896), 
and the reanalysis of Qiu et al's $W_{\lambda}$ data by Saffe \& Levato (2004). 
Unfortunately, none of these appear to be sufficiently credible as viewed from 
the present knowledge, because of neglecting the non-LTE effect for the strong
resonance D$_{1}$/D$_{2}$ lines which were invoked in all these
studies (see below in this section). 

The situation is even worse for Li and K, for which any determination of
their abundances in Vega has never been reported.
Gerbaldi, Fraggiana \& Castelli (1995) once tried to measure 
the $W_{\lambda}$(Li~{\sc i} 6708) value of Vega and derived 
a tentative value of 2\,m$\rm\AA$, 
which was not used however, since they did not regard it 
as real under the estimated ambiguity of 4\,m$\rm\AA$ (cf. their Table 6).
Regarding the K~{\sc i} 7665/7699 lines, even a trial of measuring their 
strengths has never been undertaken for Vega, as far as we know. 

Several reasons may be enumerated concerning this evident scarcity of 
investigations on Li, Na, and K (which actually holds for early-type stars 
in general):\\
--- First of all, since the strengths of Li~{\sc i}, Na~{\sc i}, and K~{\sc i} 
lines quickly fade out as $T_{\rm eff}$ becomes higher owing to the enhanced
ionization of these alkali atoms (only one valence electron being
weakly bound), one has to contend with the difficult task of measuring
very weak lines (e.g., $W_{\lambda}$ value even as small as $\sim$\,1m$\rm\AA$),
which requires a spectrum of considerably high quality.\\
--- Second, these lines situate in the longer wavelength region of
yellow--red to near-IR, where many telluric lines due to H$_{2}$O or O$_{2}$
generally exist to cause severe blending with stellar lines,
which often prevent us from reliable measurements of equivalent widths.
(see, e.g., Sect. 6.2 of Qiu et al. 2001, who remarked the large uncertainty
caused by this effect in their measurements of Na~{\sc i} D lines).\\
--- Third, the resonance lines in the longer wavelength region 
are generally known to subject to considerable non-LTE corrections
especially for the case of Na~{\sc i} 5889/5896 lines showing considerable
strengths. Unfortunately, since detailed non-LTE studies of these alkali 
elements have been rarely published as far as early-type (A--F) dwarfs are 
concerned,\footnote{Few available at present may be only two studies on neutral 
sodium lines: Takeda \& Takada-Hidai's (1994) non-LTE analysis of various 
Na~{\sc i} lines in $\alpha$ CMa (in addition to Procyon and A--F supergiants), 
and Andrievsky et al.'s (2002) non-LTE study of Na~{\sc i} D$_{1}$ and 
D$_{2}$ lines for late-B to early-F type dwarfs of $\lambda$~Boo candidates.}
lack of sufficient knowledge on non-LTE corrections may have hampered
investigating their abundances. 

Given this situation, we decided to spectroscopically establish Vega's 
photospheric abundances of Li, Na, and K scarcely investigated so far, 
while overcoming the problems mentioned above, in order to see how they are 
compared with the compositions of other elements in Vega as well as 
with those of interstellar gas. This is the purpose of this study.

This attempt was originally motivated by our recent work (Takeda, Kawanomoto 
\& Ohishi 2007) of publishing a digital atlas of Vega's high-resolution 
($R\sim 100000$) and high-S/N ($\sim$\,1000--3000) spectrum along with that of 
Regulus (a rapid rotator to be used for a reference of contaminating telluric lines), 
by which the confronted main obstacles (the necessity of measuring very weak 
lines as well as removing telluric blends) may be adequately conquered.
Besides, regarding the technical side of abundance determinations, we are 
ready to evaluate not only the necessary non-LTE corrections for Li, Na, and K 
(according to our recent experiences of statistical equilibrium calculations 
on these elements for F--G--K stars), but also the gravity-darkening 
corrections which may be appreciable especially for very weak lines as revealed 
from our recent modeling of rapidly-rotating Vega (Takeda, Kawanomoto \& Ohishi 
2008). Therefore, this would make a timely subject to address.

\section{Observational data}

The basic observational data of Vega used for this study are the high-S/N 
($\sim$~1000--3000) and high resolution ($R\sim 100000$) spectra, 
which are based on the data obtained with the HIgh Dispersion Echelle 
Spectrograph (HIDES) at the coud\'{e} focus of the 188~cm reflector of Okayama 
Astrophysical Observatory.  These spectral data of Vega, along with those of 
Regulus (a rapid rotator serving as a reference of telluric lines) were already 
published as a digital atlas by Takeda et al. (2007)\footnote{ 
While the spectrum atlas is presented as the electronic tables of this paper,
the same material is available also at the anonymous FTP site of the
Astronomical Data Center of National Astronomical Observatory of
Japan: $\langle$ftp://dbc.nao.ac.jp/DBC/ADACnew/J/other/PASJ/59.245/$\rangle$.
} which may be consulted for more details.

The following 11 lines were selected as the target lines used for
abundance determinations: Na~{\sc i} 5890/5896, 8183/8195, 6154/6161, 
5683/5688, K~{\sc i} 7665/7699, and Li~{\sc i} 6708 (cf. Table 1).
Since most of these wavelength regions (except for those of Na~{\sc i} 6154/6161 
and Li~{\sc i} 6708) are contaminated by telluric lines, these features were 
removed by dividing Vega's spectra by the reference spectra of Regulus 
(with the help of IRAF\footnote{IRAF is distributed 
by the National Optical Astronomy Observatories, which is operated by the 
Association of Universities for Research in Astronomy, Inc. under cooperative 
agreement with the National Science Foundation, USA.} task {\tt telluric}) 
which are also included in Takeda et al. (2007) along with the airmass information. 
Though this removal process did not always work very successfully (e.g., for the 
case of very strong saturated lines such as the O$_{2}$ in the K~{\sc i} 7665/7699 
region and the H$_{2}$O lines in the Na~{\sc i} 8183/8195 region), we could 
manage to recover the intrinsic stellar lines more or less satisfactorily, 
since none of these target lines fortunately coincided with the deep core 
positions of such very strong telluric features.

While most of the lines were confidently identified by comparing them with 
the theoretically simulated spectra, the detection of the extremely weak 
Li~{\sc i} 6708 line was still uncertain in the original spectrum even at
its very high S/N ratio of $\sim 2000$. Therefore, an extra smoothing
was applied to the spectrum portion in the neighborhood of this line
by taking a running mean over 9 pixels, by which the feature became
more distinct (cf. Fig.~6) thanks to the further improved S/N ratio 
by a factor of 3.\footnote{Though this results in a blurring of 
$\sim$~10~km~s$^{-1}$ corresponding to the width of 9-pixel boxcar function 
(1 pixel $\sim$ 0.03 $\rm\AA$), this does not cause any serious problem 
because it is smaller than the intrinsic line width of $\sim$~20--30 km~s$^{-1}$.} 

The measurements of the equivalent widths (with respect to the local 
continuum level specified by eye-inspection) were done by the Gaussian-fitting 
technique, and the resulting $W_{\lambda}$ values are given in Table 1. 
The relevant spectra at each of the wavelength regions are shown in 
Figs. 1 (Na\,{\sc i} 5890/5896), 2 (Na~{\sc i} 8183/8195), 
3 (Na~{\sc i} 6154/6161), 4 (Na~{\sc i} 5683/5688), 
5 (K~{\sc i} 7665/7699), and 6 (Li~{\sc i} 6708).

\begin{figure}
\includegraphics[width=9.0cm]{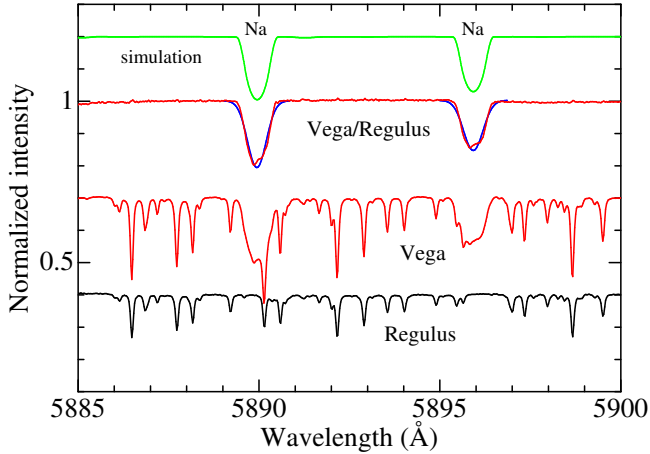}
\caption{Spectra of 5885--5900\,$\rm\AA$ region comprising
Na\,{\sc i} 5890 and 5896 lines. Theoretical simulation (LTE,
broadened with $v_{\rm e} \sin i = 22$\,km\,s$^{-1}$),
Vega/Regulus (telluric lines removed), raw Vega, and raw Regulus 
are arranged from top to bottom with appropriately chosen offsets. 
The scale of the ordinate corresponds to the second-top spectrum,
for which equivalent-widths of these lines were measured by the 
Gaussian fitting method (as overplotted in solid lines on this 
spectrum). }
\label{fig1}
\end{figure}

\begin{figure}
\includegraphics[width=9.0cm]{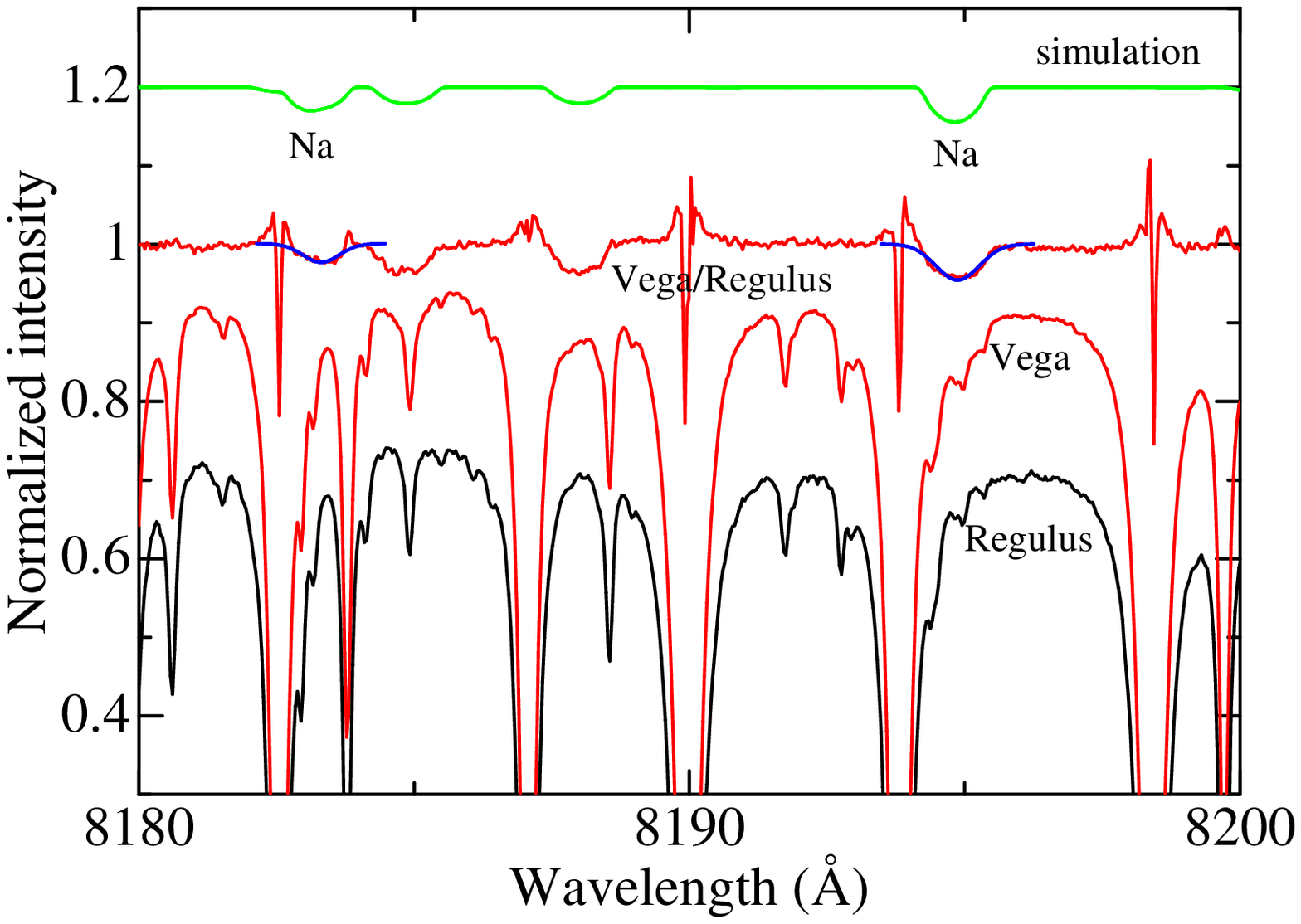}
\caption{Spectra of 8180--8200\,$\rm\AA$ region comprising
Na\,{\sc i} 8183 and 8195 lines. Theoretical simulation,
Vega/Regulus (telluric lines removed), raw Vega, and raw Regulus 
are arranged from top to bottom with appropriately chosen offsets.
Otherwise, the same as in Fig.~1.}
\label{fig2}
\end{figure}

\begin{figure}
\includegraphics[width=9.0cm]{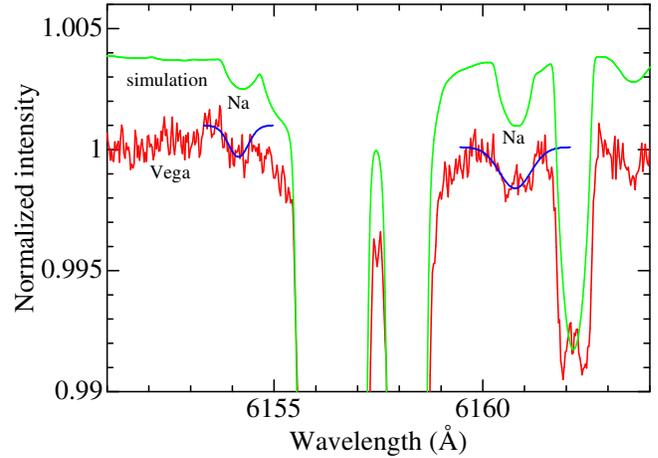}
\caption{Spectra of 6151--6164\,$\rm\AA$ region comprising
Na\,{\sc i} 6154 and 6161 lines. Theoretical simulation (upper)
and raw Vega (lower) are placed with an appropriate offset.
This region is not contaminated by telluric lines.
Otherwise, the same as in Fig.~1.}
\label{fig3}
\end{figure}

\begin{figure}
\includegraphics[width=9.0cm]{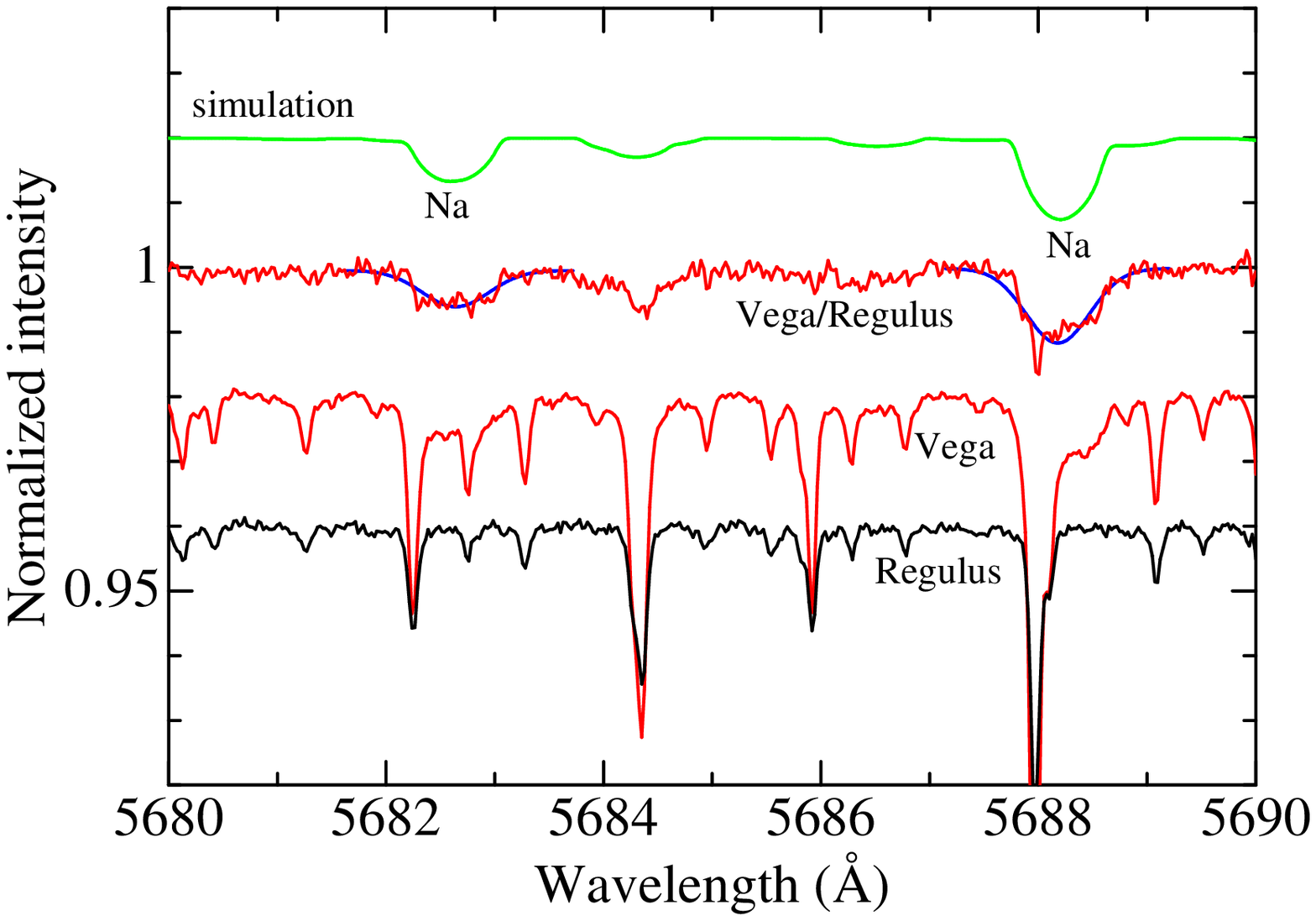}
\caption{Spectra of 5680--5690\,$\rm\AA$ region comprising
Na\,{\sc i} 5683 and 5688 lines. Theoretical simulation,
Vega/Regulus (telluric lines removed), raw Vega, and raw Regulus 
are arranged from top to bottom with appropriately chosen offsets.
Otherwise, the same as in Fig.~1.}
\label{fig4}
\end{figure}

\begin{figure}
\includegraphics[width=9.0cm]{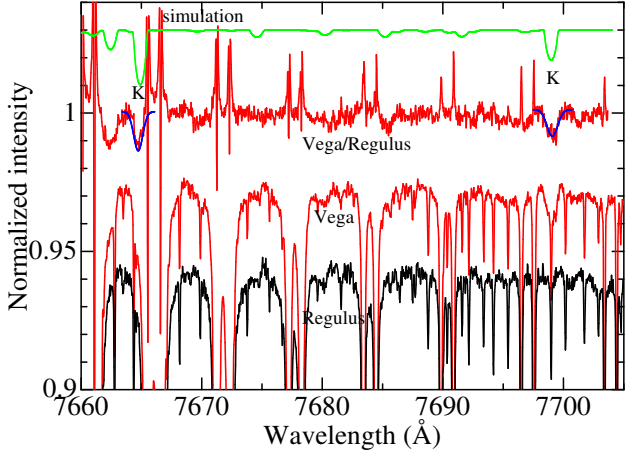}
\caption{Spectra of 7660--7705\,$\rm\AA$ region comprising
K\,{\sc i} 7665 and 7699 lines. Theoretical simulation,
Vega/Regulus (telluric lines removed), raw Vega, and raw Regulus 
are arranged from top to bottom with appropriately chosen offsets.
Otherwise, the same as in Fig.~1.}
\label{fig5}
\end{figure}

\begin{figure}
\includegraphics[width=9.0cm]{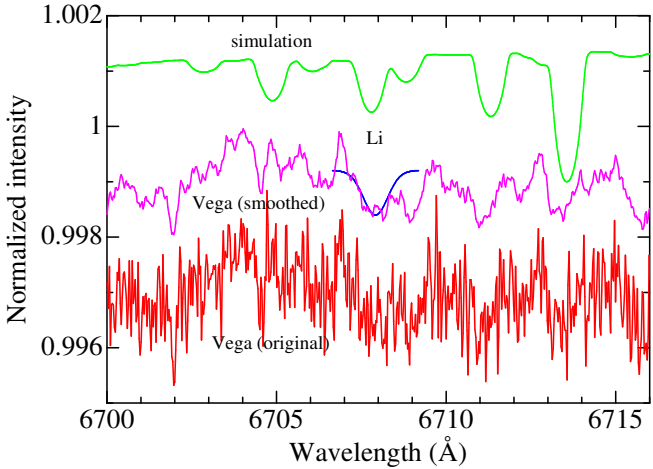}
\caption{Spectra of 6700--6716\,$\rm\AA$ region comprising
Li\,{\sc i} 6708 merged doublet. Theoretical simulation, smoothed Vega spectrum 
by convolving the 9-pixel boxcar function, and raw Vega spectrum 
are arranged from top to bottom with appropriately chosen offsets.
This region is hardly contaminated by telluric lines.
Otherwise, the same as in Fig.~1.}
\label{fig6}
\end{figure}

\section{Abundance determinations}

Regarding the fiducial atmospheric model of Vega, Kurucz's (1993) ATLAS9 model 
with the parameters of $T_{\rm eff} = 9550$~K (effective temperature), 
$\log g$ (cm~s$^{-2}$) = 3.95 (surface gravity), [X/H] = $-0.5$ (metallicity), 
and $v_{\rm t}$ = 2~km~s$^{-1}$ (microturbulence) was adopted for this study 
as in Takeda et al. (2007), which appears to be the best as far as within 
the framework of classical plane-parallel ATLAS9 models according to 
Castelli \& Kurucz (1994).

As the basic strategy, the LTE abundance ($A_{\rm LTE}$) was first derived
from the measured $W_{\lambda}$ by using the WIDTH9 program,\footnote{
This is a companion program to the ATLAS9 model atmosphere program written 
by R. L. Kurucz (Kurucz 1993), though it has been considerably modified by 
Y. T. (e.g., for treating multiple component lines or including non-LTE effects 
in line formation}, to which the non-LTE correction ($\Delta_{\rm NLTE}$) and 
the gravity-darkening correction ($\Delta_{\rm GD}$) were then applied to 
obtain the final abundance ($A_{\rm NLTE+GD}$). 

In this step-by-step way,\footnote{Complete simulations (where the 
non-LTE effect and the gravity-darkening effect were simultaneously 
taken into account in a rigorous manner) show that the finally derived 
abundances in this semi-classical approach are fairly reliable. 
This consistency check is separately described in Appendix B.}
one can quantitatively judge the importance/unimportance 
of each effect by comparing two abundance corrections, and
such factorized information may be useful in application to
other different situations (e.g., the case of slow rotator where only 
$\Delta_{\rm LTE}$ is relevant; cf. Appendix A).
Also, the use of ``relative correction'' (instead of directly
approaching the absolute abundance solution)
has a practical merit of circumventing the numerical-precision problem
involved in the spectrum computation on the gravity-darkened model 
(especially seen for the case of the extremely weak Li~{\sc i} 6708 line
composed of several components; cf. footnote 9). 

The non-LTE calculations for evaluating $\Delta_{\rm NLTE}$ were carried out 
in the same manner as described in Takeda et al. (2003; Na~{\sc i}), 
Takeda \& Kawanomoto (2005; Li~{\sc i}), and Takeda et al. (2002; K~{\sc i}).
Besides, all the relevant atomic data ($gf$ values, damping constants, etc.) 
for abundance determinations were so adopted as to be consistent with 
these three studies.

The gravity-darkening corrections ($\Delta_{\rm GD}$) were derived based on
the equivalent-width intensification factor computed by the program
CALSPEC under the assumption of LTE ($W_{4}/W_{0}$, where $W_{4}$ 
and $W_{0}$ correspond to the gravity-darkened rapid rotator model and 
the classical rigid-rotation model, respectively; and the abundance
was adjusted to make $W_{4}$ consistent with $W_{\lambda}^{\rm obs}$), 
the details about which are explained in Takeda et al. (2008). 
Practically, the relation 
$\Delta_{\rm GD} = -\log (W_{4}/W_{0})$ was assumed for the very weak lines
($W_{\lambda} < 15$~m$\rm\AA$; i.e., Na~{\sc i} 6154/6161, Na~{\sc i} 5683/5688, 
K~{\sc i} 7665/7699, and Li~{\sc i} 6708) being guaranteed to locate on the
linear part of the curve of growth. The $\Delta_{\rm GD}$ values for the 
remaining four lines (Na~{\sc i} 5890/5896 and Na~{\sc i} 8183/8195) 
were directly obtained from the abundance difference 
between those derived from the true $W_{\lambda}^{\rm obs}$ and the 
perturbed $W_{\lambda}^{\rm obs}/(W_{4}/W_{0})$. 

The results of the abundances and the abundance corrections are summarized
in Table 1, where the finally adopted (average) values of 
$\langle A_{\rm NLTE+GD} \rangle$ and the abundances relative to the
standard solar system compositions 
([X/H] $\equiv \langle A_{\rm NLTE+GD}^{\rm X} \rangle- A^{\odot}$) 
are also presented. As can be seen from this table, the non-LTE corrections
are always negative (corresponding to the non-LTE line-intensification) 
with appreciably $W_{\lambda}$-dependent extents (ranging from 0.1--0.7 dex).
Similarly, the gravity-darkening corrections are also negative with extents 
of 0.1--0.3 dex\footnote{It may be worth noting that these moderate 
$\Delta_{\rm GD}$ values correspond to Takeda et al.'s (2008) model No. 4 
($v_{\rm e} = 175$~km~s$^{-1}$) which they concluded to be the best.
If higher-rotating models with larger gravity-darkening
were to be used, the corrections would naturally become more enhanced.
For example, if we adopt Takeda et al.'s (2008) model No. 8
with $v_{\rm e} = 275$~km~s$^{-1}$, which is near to the solution
suggested from interferometric observations (Peterson et al. 2006,
Aufdenberg et al. 2006), the extents of $\Delta_{\rm GD}$ turn out
to increase significantly by a factor of $\sim 3$ (i.e., $-0.22$, $-0.18$, 
$-0.31$, $-0.23$, $-0.42$, $-0.43$, $-0.36$, $-0.32$, $-0.34$, $-0.37$,
and $-0.82$, for Na~{\sc i} 5890, 5896, 8183, 8195, 6154, 6161, 5683, 5688,
K~{\sc i} 7664, 7699, and Li~{\sc i} 6708, respectively).}
(especially important for the weakest Li~{\sc i} 6708 line) 
reflecting the line-strengthening caused by rotation-induced $T$-lowering.

Regarding errors in the resulting abundances, those due to uncertainties 
in the atmospheric parameters (e.g., in $T_{\rm eff}$ or $\log g$) may  
not be very significant in the present comparatively well-established
case (an abundance change is only $\sim 0.1$~dex for a change either 
in $T_{\rm eff}$ by $\sim 200$~K or in $\log g$ by 0.2; cf. Table 5 in 
Takeda \& Takada-Hidai 1994 for the case of $\alpha$ CMa).
Rather, a more important source of error would be the ambiguities in 
the measurement of $W_{\lambda}$. While the photometric accuracy of
$W_{\lambda}$ estimated by applying Cayrel's (1988) formula 
(FWHM of $w \sim 0.5\rm\AA$, pixel size of $\delta x \sim 0.03 \rm\AA$, 
and photometric error of $\epsilon \sim 1/2000$) is only on the order of 
$\sim 0.1$~m$\rm\AA$ and inconsiderable, uncertainties in specifying the 
continuum level can be a more serious problem, especially for the case of 
extremely weak lines where the relative depression is only on the order of 
$\sim 10^{-3}$. Presumably, errors in $W_{\lambda}$ would amount to 
several tens of \% in the case of very faint lines (e.g., Na~{\sc i} 6154 
or Li~{\sc i} 6708) or the case of insufficient spectrum quality due to 
incompletely removed telluric lines (e.g., Na~{\sc i} 8183 or K~{\sc i} 7665). 
Accordingly, the abundance results derived from such problematic lines 
may be uncertain by up to $\sim~$0.2 dex, which should particularly be 
kept in mind in discussing the final abundances of Li and K being based on 
only 1 or 2 lines.

\section{Results and Discussions}

\subsection{Na abundance of Vega}

The abundances of sodium could be derived from 8 lines of different strengths 
(2 resonance lines of D$_{1}$+D$_{2}$ and 6 subordinate lines originating 
from the excited level of $\chi_{\rm low} = 2.21$~eV).
The three kinds of abundances ($A_{\rm LTE}$, $A_{\rm NLTE}$, $A_{\rm NLTE+GD}$; 
cf. Table 1) for each of the 8 lines are plotted against $W_{\lambda}$ in 
Fig.~7. We can read from this figure that the serious inconsistency
seen in the LTE abundances is satisfactorily removed by the non-LTE
corrections, which are significantly $W_{\lambda}$-dependent from 
$-0.7$ dex (for the strongest Na~{\sc i} 5890) to $-0.1$ dex (for the
weakest-class Na~{\sc i} 6154/6161/5683/5688 lines of $W_{\lambda} < 10$~m$\rm\AA$).
This may suggest that the applied non-LTE corrections are fairly 
reliable and the adopted microturbulence of $v_{\rm t}$ = 2~km~s$^{-1}$ 
is adequate (which affects the abundances of stronger Na~{\sc i} 5890/5896 
lines; cf. Table 5 of Takeda \& Takada-Hidai 1994).
The final sodium abundance of Vega, obtained by averaging $A_{\rm NLTE+GD}$ 
values of 8 lines, is 6.01 with the standard deviation of $\sigma = 0.10$, 
indicating a mildly subsolar composition ([Na/H] = $-0.3$) compared to
the solar abundance of $A_{\odot} = 6.31$.

As mentioned in Sect. 1, a few previous determinations based on D$_{1}$+D$_{2}$ 
lines are available for the Na abundance of Vega. Hunger (1955) tentatively 
derived $A \sim 6.7$ (coarse analysis) and $A \sim 7.8$ (fine analysis) 
from $W_{\lambda}$(5890/5896) = 195/115 m$\rm\AA$, though commenting 
that these abundances are very uncertain. Thereafter, based on 
Hunger's $W_{\lambda}$ data, Strom et al.'s (1966) model atmosphere analysis 
concluded $A \sim 7.7 (\pm 0.4)$ ($T_{\rm eff} = 9500$~K).
Then, after a long blank, Qiu et al. (2001) obtained $A = 6.45$ from
their measured $W_{\lambda}$(5896) value of 94~m$\rm\AA$, though they also
remarked that this result is quite uncertain because of blending with
water vapor lines. Soon after, Saffe \& Levato (2004) derived $A = 6.37$
using Qiu et al.'s $W_{\lambda}$ data. Accordingly, all these previous
work reported supersolar or near-solar Na abundances for Vega, contradicting
the conclusion of this study. It is evident that they failed to
obtain the correct abundance from the Na~{\sc i} 5890/5896 lines
because of neglecting the non-LTE effect, which is substantially important
for these strong resonance lines.

\begin{figure}
\includegraphics[width=9.0cm]{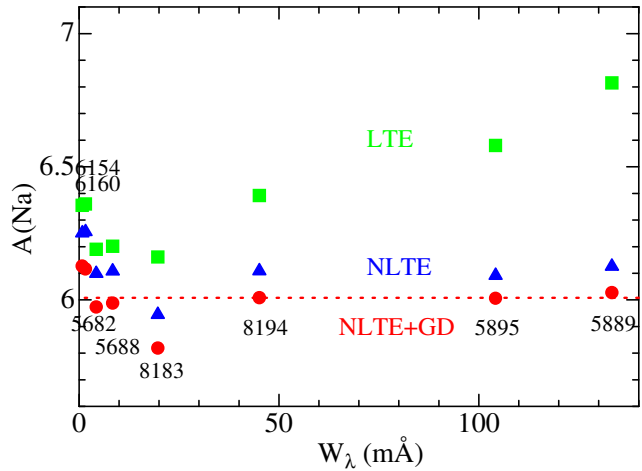}
\caption{Na abundances derived from each of the 8 lines
plotted against the equivalent widths, based on the results
in Table 1. Squares, triangles, and circles correspond to
the LTE abundances, the non-LTE abundances, and the final abundances
(including both the non-LTE and gravity-darkening corrections),
respectively. The horizontal dotted line indicates the average
of the final abundances (6.01).}
\label{fig7}
\end{figure}

\subsection{Composition characteristics of Li, Na, and K}

According to Table 1, while potassium is only mildly subsolar ([K/H] 
$\simeq -0.2$) to an extent similar to sodium ($\sim -0.3$), lithium is 
more manifestly underabundant as [Li/H] $\simeq -0.6$. Therefore, we may 
conclude that the deficiency in Li is markedly different from that of Na 
and K, even considering the possible ambiguities of $\sim 0.2$\,dex 
(cf. Sect. 3). How should we interpret these results? Do they have something 
to do with the element depletion (due to dust condensation) in interstellar 
gas and its accretion? Since the quantitative amount of such a depletion 
is different from case to case depending on the physical condition
(e.g., see the variety of [Na/H]$_{\rm ISM}$ in Fig.~7 of Paunzen et al. 
2002), it is not much meaningful to discuss the ``absolute'' values of [X/H] 
here. Rather, we should pay attention to their ``relative'' behaviors 
with each other or in comparison to those of various other elements.

The [X/H]$_{\rm Vega}$ values (those for Li, Na, and K are from this study,
while those for other elements were taken from various literature)
are plotted against the condensation temperature ($T_{\rm c}$) in Fig.~8 
(filled circles), where the [X/H]$_{\rm ISM}$ results of typical interstellar 
gas in the direction of $\zeta$ Oph (taken from Table 5 of Savage \& Sembach 1996; 
see also Fig.~4 therein) are also shown by open circles.
The following characteristics can be recognized from this figure:\\
--- (1) The [X/H]$_{\rm Vega}$ values, falling on a rather narrow range between 
$\sim -1$ and $\sim 0$, tend to decrease with $T_{\rm c}$, a qualitatively
similar trend to that seen in [X/H]$_{\rm ISM}$.\\  
--- (2) The run of [X/H]$_{\rm Vega}$ with $T_{\rm c}$ appears to be slightly 
discontinuous at $T_{\rm c} \sim 1000$\,K 
(i.e., [X/H]$_{\rm Vega} \sim -0.2$ to $-0.3$ at $T_{\rm c} \la 1000$\,K and
[X/H]$_{\rm Vega} \sim -0.5$ to $-0.6$ at $T_{\rm c} \ga 1000$\,K).\\
--- (3) Interestingly, the decreasing [X/H]$_{\rm ISM}$ with $T_{\rm c}$ also 
shows a discontinuity around this critical $T_{\rm c}$ ($\sim 1000$\,K) 
in the sense that the slope of $|d$[X/H]$_{\rm ISM}/dT_{\rm c}|$
becomes evidently steeper on the $T_{\rm c} \ga 1000$\,K side.\\
--- (4) Specifically, the tendency of [Li/H]$_{\rm Vega}$ ($-0.6$) being 
more deficient than [Na/H]$_{\rm Vega}$ ($-0.3$) and [K/H]$_{\rm Vega}$ 
($-0.2$) is qualitatively similar to just what is seen in ISM 
([Li/H]$_{\rm ISM}<$ [Na/H]$_{\rm ISM} \simeq$ [K/H]$_{\rm ISM}$).

All these observational facts may suggest the existence of some kind of 
connection between Vega's photospheric abundances and those of interstellar 
cloud, which naturally implies that an accretion/contamination of interstellar 
gas is likely to be responsible (at least partly) for the abundance 
peculiarity of Vega, such as being proposed for explaining the $\lambda$ Boo 
phenomenon. This is the conclusion of this study.

\begin{figure}
\includegraphics[width=9.0cm]{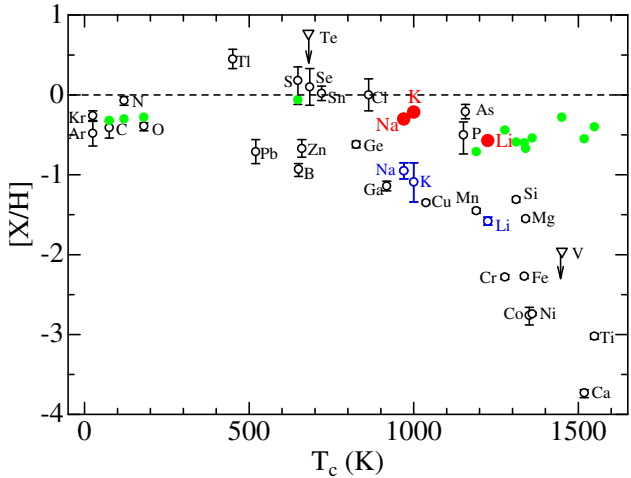}
\caption{[X/H] values (relative abundances in comparison with
the solar-system values of Grevesse \& Noels 1993) of various elements 
plotted against the condensation temperature ($T_{\rm c}$). 
Open circles indicate the compositions of interstellar gas in the 
direction of $\zeta$~Oph (cool diffuse clouds), which were taken from
Savage \& Sembach (1996; cf. their Table 5 and Fig.~4). 
The [Na/H], [K/H], and [Li/H] for Vega derived in this study
are shown by larger filled circles. Besides, Vega's [X/H] values
for other elements are plotted by smaller filled circles for a reference, 
which were taken from from various sources:  Przybilla \& Butler 
(2001) (for C, N, and O; cf. their Table 7); Takada-Hidai \& Takeda (1996) 
(for S); Adelman \& Gulliver (1990) (for Mg, Ca, Ti, Cr, Mn, Fe, and Ni);
Qiu et al. (2001) (for Si and V).
[Note. Since these literature [X/H] values of Vega were derived in a 
conventional manner based on classical atmospheric model, further 
corrections for the gravity-darkening effect ($\Delta_{\rm GD}$) are 
to be expected. 
According to Takeda et al. (2008), however, the extents of (negative) 
$\Delta_{\rm GD}$ (becoming appreciable only for very weak lines and 
for specific elements/stages) are only $\sim 0.2$~dex at most (e.g., 
Adelman \& Gulliver's [Ca/H] value derived from very weak Ca~{\sc i} 
lines had better be reduced by $\sim 0.2$~dex) while $\la 0.1$~dex 
in many typical cases, which are thus unlikely to cause any significant 
change in the general pattern of [X/H] shown here.]
}
\label{fig8}
\end{figure}

\section*{Acknowledgments}

The author thanks S. Kawanomoto and N. Ohishi for their help
in the observations of Vega, based on the data from which 
this study is based.

\setcounter{table}{0}
\begin{table*}
\begin{minipage}{130mm}
\small
\caption{Atomic data, equivalent widths, and abundance results.}
\begin{center}
\begin{tabular}{cccccrccccc}\hline
Species & RMT & $\lambda$ & $\chi_{\rm low}$ & $\log gf$ & $W_{\lambda}$ & $A_{\rm LTE}$ & 
$\Delta_{\rm NLTE}$ & $\Delta_{\rm GD}$ & $A_{\rm NLTE+GD}$ & [X/H] \\ 
 &  &  ($\rm\AA$) & (eV) &  &  (m$\rm\AA$) &  &  &  &  &  \\
\hline
Na\,{\sc i} & 1  & 5889.95  & 0.00  & +0.12  & 133.3  & 6.82  & $-$0.69  & $-$0.09  & 6.03  &  \\ 
Na\,{\sc i} & 1  & 5895.92  & 0.00  & $-$0.18  & 104.2  & 6.58  & $-$0.49  & $-$0.08  & 6.01  &  \\
Na\,{\sc i} & 4  & 8183.26  & 2.10  & +0.22  &  19.7  & 6.16  & $-$0.22  & $-$0.12  & 5.82  &  \\
Na\,{\sc i} & 4  & 8194.82  & 2.10  & +0.52  &  45.1  & 6.39  & $-$0.28  & $-$0.10  & 6.01  &  \\
Na\,{\sc i} & 5  & 6154.23  & 2.10  & $-$1.56  &   0.8  & 6.36  & $-$0.10  & $-$0.12  & 6.13  &  \\
Na\,{\sc i} & 5  & 6160.75  & 2.10  & $-$1.26  &   1.6  & 6.36  & $-$0.10  & $-$0.14  & 6.12  &  \\
Na\,{\sc i} & 6  & 5682.63  & 2.10  & $-$0.67  &   4.3  & 6.19  & $-$0.09  & $-$0.13  & 5.97  &  \\
Na\,{\sc i} & 6  & 5688.21  & 2.10  & $-$0.37  &   8.4  & 6.20  & $-$0.09  & $-$0.12  & 5.99  &  \\
 &  &  &  &  &  &  &  &  & {\it 6.01}  & {\it $-$0.30}  \\ 
\hline
K\,{\sc i} & 1  & 7664.91  & 0.00  & +0.13  & 13.3  & 5.23  & $-$0.28  & $-$0.13  & 4.82  \\ 
K\,{\sc i} & 1  & 7698.97  & 0.00  & $-$0.17  & 10.7  & 5.42  & $-$0.27  & $-$0.14  & 5.01  \\ 
 &  &  &  &  &  &  &  &  & {\it 4.91}  & {\it $-$0.22} \\
\hline
Li\,{\sc i} & 1  & 6707.756  & 0.00  & $-$0.43  & 0.7  & 3.15  & $-$0.15  & $-$0.26  & 2.74  &  \\ 
 &  & 6707.768  &  & $-$0.21  &  &  &  &  &  &  \\ 
 &  & 6707.907  &  & $-$0.93  &  &  &  &  &  &  \\
 &  & 6707.908  &  & $-$1.16  &  &  &  &  &  &  \\
 &  & 6707.919  &  & $-$0.71  &  &  &  &  &  &  \\
 &  & 6707.920  &  & $-$0.93  &  &  &  &  &  &  \\ 
 &  &  &  &  &  &  &  &  & {\it 2.74}  & {\it $-$0.57} \\
\hline
\end{tabular}
\end{center}
Following the atomic data of spectral lines (species, multiplet No., wavelength,
lower excitation potential, and logarithmic $gf$ value) in columns 1--5,
column 6 gives the measured equivalent width. The results of the abundance analysis are 
given in column 7 (LTE abundance; in the usual normalization of H=12.00),
column 8 (non-LTE correction), column 9 (gravity-darkening correction),
and column 10 (non-LTE as well as gravity-darkening corrected abundance).  
The finally adopted average abundance and the corresponding [X/H] value 
($\equiv A_{\rm X}^{\rm Vega} - A_{\rm X}^{\odot}$) are also given at the bottom of 
the section (in italics; columns 10 and 11),
where Grevesse \& Noels's (1993) values of 6.31 (Na), 3.31 (Li), and 5.13 (K) were used 
as the standard solar-system abundances ($A_{\rm X}^{\odot}$).
Note that the analysis of $W_{\lambda}$(Li\,{\sc i}~6708) was done by synthesizing the six 
component lines of $^{7}$Li (neglecting the contribution of $^{6}$Li; cf. Takeda \& 
Kawanomoto 2005), 
while all the remaining lines of Na and K were analyzed by the single-line treatment
as was done by Takeda et al. (2003; Na) and Takeda et al. (2002; K).
\end{minipage}
\end{table*}

\appendix

\section{Li abundance in $o$~Peg}

The enhanced deficiency of Li (compared to Na and K) was an important 
key result in deriving the conclusion of this study (cf. Sect. 4.2),
since we interpreted it as a manifestation of ISM compositions.
In this connection, it may be worth mentioning another remarkable 
very sharp-lined A1~IV star, $o$~Peg, for which we could also get 
information of the Li abundance. 
As the only available measurement of the Li~{\sc i} 6708 line for early 
A-type stars thus far to our knowledge, Coupry \& Burkhart (1992) derived 
$W_{\lambda}$(6708) = 1.3~m$\rm\AA$ for this star, which has atmospheric 
parameters ($T_{\rm eff}$ = 9650~K, $\log g = 3.6$) quite similar to 
those of Vega except for its near-normal metallicity ([Fe/H] 
$\simeq +0.1$; Burkhart \& Coupry 1991). 

Now, their $W_{\lambda}$($o$~Peg) value twice as large as that of Vega 
(0.7~m$\rm\AA$) would raise the Li abundance by +0.3 dex. Furthermore, since 
any flat-bottomed shape has never been reported in the spectral lines 
of $o$~Peg (as we can confirm from the high-quality spectrum atlas\footnote{
Also available at \\
http://www.brandonu.ca/physics/gulliver/ccd\_atlases.html .}
of this star published by Gulliver, Adelman, \& Friesen 2004), the 
application of the gravity-darkening correction ($\sim -0.3$\,dex for Vega) 
may not be relevant here, which again leads to an increase of 0.3\,dex. 
Hence, we have [Li/H]$_{\rm o Peg} (\simeq$~[Li/H]$_{\rm Vega}$ + 0.6) 
$\sim 0.0$; an interesting result that the photospheric Li abundance of 
$o$~Peg almost coincides with that of the solar-system composition.
This result, that [Li/H] (as well as [Fe/H]) is deficient/normal in 
Vega/$o$~Peg, may suggest that the mechanism responsible for producing the 
underabundance of these elements in Vega is irrelevant for $o$~Peg. Thus, 
according to our interpretation, $o$~Peg would not have suffered any 
pollution due to accretion of interstellar gas depleted in volatile
elements of higher $T_{\rm c}$. 

Yet, we had better keep in mind another possibility that the mechanism 
causing this Li deficiency in Vega might different from that of other metals.
For example, this underabundance could be attributed to some process of 
envelope mixing (e.g., meridional circulation or shear-induced 
turbulence which are supposed to be more significant as a star rotates
faster), because Li atoms are burned and destroyed when they are 
conveyed into the hot stellar interior ($T \ga 2.4 \times 10^{6}$\,K).
Namely, since we know that Vega rotates rapidly (as fast as 
$v_{\rm e} \sim 200$~km~s$^{-1}$) while $o$~Peg does not so much 
(at least the gravity-darkening effect is not so significant as in Vega), 
the underabundance of Li in the atmosphere of Vega might stem from the 
rotation-induced mixing (which is not expected for slowly rotating $o$~Peg).
If this is the case, however, the rough similarity in the extent of 
deficiency in Li as well as other metals has to be regarded as a mere
coincidence, which makes us feel this possibility as rather unlikely.

\section{Check for the final abundances: Complete spectrum synthesis}

Our basic strategy for deriving the abundances of Na, K, and Li in 
Sect. 3 was as follows. \\
--- (1) First, the $A_{\rm LTE}$ was derived from 
$W_{\lambda}^{\rm obs}$ based on the classical plane-parallel model 
atmosphere. \\
--- (2) Next, the non-LTE correction ($\Delta_{\rm NLTE}$) was 
derived in the conventional way by using this classical model. \\
--- (3) Then, the gravity darkening correction ($\Delta_{\rm GD}$)
was evaluated from the line-intensification factor 
$W_{4}^{\rm LTE}/W_{0}^{\rm LTE}$ (where the abundance was so adjusted
as to satisfy $W_{4}^{\rm LTE} \simeq W_{\lambda}^{\rm obs}$), which 
was computed by applying the CALSPEC program (Takeda et al. 2008) 
with the assumption of LTE to the gravity-darkened model~4 and 
the rigid-rotation model~0. \\
--- (4) Finally, $A_{\rm NLTE+GD}$ was
obtained as $A_{\rm LTE} + \Delta_{\rm NLTE} + \Delta_{\rm GD}$. 

Actually, such a phased approach has a distinct merit 
of clarifying the importance/contribution of two different effects 
(the non-LTE effect and the gravity-darkening effect).
Besides, from a practical point of view, the necessary amount of 
calculations to arrive at the final abundance solution (such that 
reproducing the observed spectrum) can be considerably saved.

However, there is some concern about whether such a step-by-step approach
(treating the two effects separately) really yield sufficiently correct 
results, because both are actually related with each other (e.g., how 
does the largely variable non-LTE corrections over the gravity-darkened 
stellar surface differing in $T$ or $g$ play roles? How reliable is the 
gravity-darkening correction derived by neglecting the non-LTE effect?). 
Therefore, it may be worth checking the validity of the finally derived 
abundances ($A_{\rm NLTE+GD}$) by carrying out
a complete spectrum synthesis including both NLTE and GD effects 
simultaneously. 
For this purpose, the CALSPEC program was modified
so as to allow inclusion of the non-LTE departure coefficients
(corresponding to the different conditions at each of the points 
over the gravity-darkened stellar surface), and the NLTE+GD
profiles ($R_{4}^{\rm NLTE}$) were computed for each of the 9 lines 
(Na~{\sc i} 5890, 5896, 8183, 8195, 6154, 6161, 5683, 5688,
K~{\sc i} 7664, 7699, and Li~{\sc i} 6708) by using the already known
$A_{\rm NLTE+GD}$ values (6.03, 6.01, 5.82, 6.01, 6.13, 6.12, 5.97,
5.99, 4.82, 5.01, and 2.74, respectively). 
Such obtained $R_{4}^{\rm NLTE}$ profiles are depicted (in thick solid 
lines) in Fig. 9, where the relevant three kinds of line profiles 
[$R_{0}^{\rm NLTE}$ (thick dashed line), $R_{4}^{\rm LTE}$ (thin solid 
line), and $R_{0}^{\rm LTE}$ (thin dashed line)] are also shown for 
comparison.

\begin{figure}
\includegraphics[width=8.0cm]{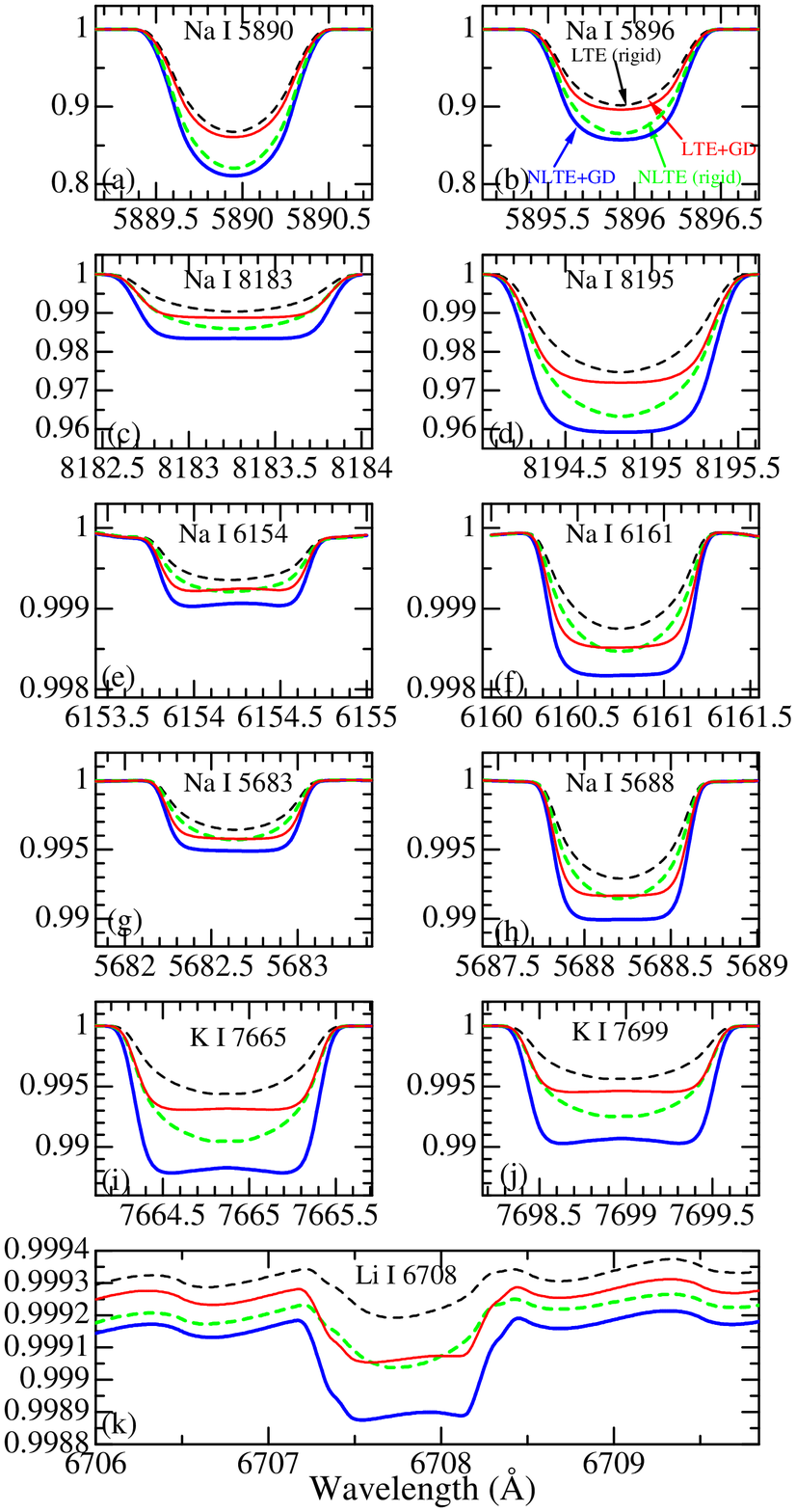}
\caption{
Theoretical profiles of the Na, K, and Li lines computed 
by the CALSPEC program (developed for synthesizing the flux spectrum 
for a given rotationally-distorted gravity-darkened stellar model; 
cf. Takeda et al. 2008), which has been modified to allow inclusion 
of the non-LTE effect. The $A_{\rm NLTE+GD}$ values given in Table 1 
were assumed as the abundances for each of the lines.
Results for the gravity-darkened model and the classical 
rigid-rotation model (model No. 4 and No. 0 in Takeda et al. 2008)
are discriminated by the line type (solid and dashed lines, 
respectively), and those for NLTE and LTE are by the line thickness
(thick and thin lines, respectively). 
As a result, four profiles are shown for each of the 9 lines: 
$R_{4}^{\rm NLTE}$ (thick solid line)
$R_{0}^{\rm NLTE}$ (thick dashed line), 
$R_{4}^{\rm LTE}$ (thin solid line), and 
$R_{0}^{\rm LTE}$ (thin dashed line).
Shown in the ordinate is the normalized flux divided by the theoretical 
(pure) continuum; therefore, the local continuum level sometimes 
turns out to be slightly less than unity because of the extended wings 
of H lines. (Note that the theoretical equivalent widths 
$W_{4}^{\rm NLTE}$ corresponding to $R_{4}^{\rm NLTE}$ profiles 
discussed in Appendix B were calculated with respect to the local continuum, 
i.e. the maximum level in the neighborhood of the line profiles, 
irrespective of the scales in the ordinate.)
(a) Na~{\sc i} 5890, 
(b) Na~{\sc i} 5896, (c) Na~{\sc i} 8183, (d) Na~{\sc i} 8195,
(e) Na~{\sc i} 6154, (f) Na~{\sc i} 6161, (g) Na~{\sc i} 5683,
(h) Na~{\sc i} 5688, (i) K~{\sc i} 7665, (j) K~{\sc i} 7699, and
(k) Li~{\sc i} 6708. }
\label{figB1}
\end{figure}

The resulting theoretical equivalent widths ($W_{4}^{\rm NLTE}$) 
computed by integrating $R_{4}^{\rm NLTE}(\lambda)$ are
129.7, 101.3, 18.4, 43.2, 0.8, 1.5, 4.0, 7.8, 12.8, 10.3, and 0.4 m$\rm\AA$
for these 9 lines, respectively. Comparing these values with the
observed equivalent widths ($W_{\lambda}^{\rm obs}$) given in Table 1,
we can see a satisfactory agreement between $W_{4}^{\rm NLTE}$ and
$W_{\lambda}^{\rm obs}$ (typically to within several percent except for 
Li~{\sc i} 6708\footnote{We see a rather large discrepancy for this 
Li line ($W_{4}^{\rm NLTE}$ = 0.4~m$\rm\AA$, while $W_{\lambda}^{\rm obs}$ 
= 0.7~m$\rm\AA$). This must be due to the inevitable numerical 
errors in $R_{4}^{\rm NLTE}$ (Li~{\sc i} 6708) computed by the CALSPEC 
program (synthesizing the flux spectrum by integrating the contributions 
from a number of finite surface elements; each $1^{\circ} \times 1^{\circ}$ 
in latitude and longitude), which becomes appreciable especially for this case of 
Li~{\sc i} 6708 line involving synthesis of several component lines. 
As a matter of fact, we can see a spurious wavy pattern in the continuum
level outside of this line (see Fig. B1k) which indicates the existence
of numerical problems (even if small). It may be worth stressing that 
the gravity-darkening correction $\Delta_{\rm GD} \simeq -\log (W_{4}/W_{0})$ 
is not affected by this numerical error in $W$ because it is canceled
by taking the ratio of $W$. In this sense, our semi-classical approach of 
establishing the final abundance (application of two corrections to the 
classical LTE abundance solution; cf. Sect. 3) is surely advantageous 
as far as this Li~{\sc i} 6708 line is concerned.}), by which we may conclude 
that the approach adopted in Sect. 3 is practically validated.

\end{document}